\documentclass{sf2a-conf2024}
\usepackage{graphicx}
\usepackage{hyperref}
\usepackage[]{natbib}  
\usepackage{epstopdf}

\def\BibTeX{{\rm B\kern-.05em{\sc i\kern-.025em b}\kern-.08em
    T\kern-.1667em\lower.7ex\hbox{E}\kern-.125emX}}
\bibpunct{(}{)}{;}{a}{}{,}  

\newcommand{\tCO}{{tCO$_2$eq}}
\newcommand{\mpa}{mag/arcsec$^2$}

\begin{document}

\TitreGlobal{SF2A 2024}


\title{Action Research in Astronomy and Ecology: The observatory of the night environment in Grenoble}
\runningtitle{The observatory of the night environment in Grenoble}

\author{J. Milli}\address{Univ. Grenoble Alpes, CNRS, IPAG, F-38000 Grenoble, France}
\author{M. Boribon$^1$} 
\author{F. Malbet$^1$}
\author{P. Deverchere} \address{ScotopicLabs, 11 rue Calas, 69004, Lyon, France}           
\author{B. Drillat} \address{LPO Auvergne Rh\^one-Alpes, d\'el\'egation territoriale de l’Is\`ere, MNEI, 5 place bir Hakeim, 38000 Grenoble, France}           
\author{B. Falque}  \address{AurorAlpes, 2 rue Anthoard, 38000 Grenoble}
\author{F. Colas}  \address{IMCCE, Observatoire de Paris, PSL Research University, CNRS UMR 8028, 77 av. Denfert-Rochereau, 75014 Paris, France}
\author{A. Malgoyre}  \address{Service Informatique Pythéas (SIP) CNRS – OSU Institut Pythéas
– UMS 3470, Marseille, France
}




\setcounter{page}{237}


\maketitle


\begin{abstract}
We present an example of low-carbon research activity carried out by astrophysicists and focused on ecology and environmental protection, with direct impacts on territories and society. This project serves as an illustration of action research for an astrophysics lab in the context of the current ecological crisis.
\end{abstract}



\section{Introduction}

\subsection{Context: astronomy in the current bioclimatic crisis}

The combined crisis of climate change and biodiversity loss represent major risks for our societies and the Earth's habitability. The responsibility of human activity is indisputable, as stated in the  sixth assessment Report of the United Nations Intergovernmental Panel on Climate Change. To respect the 2015 Paris Agreement and reach carbon neutrality by 2050, it is necessary today to cut emissions of greenhouse gases by 7\% per year. Astronomers have a large carbon footprint evaluated to $36.6 \pm 14.0$  \tCO{} per year per astronomer \citep{Knoedlseder2022}, which place them among the largest emitters of the scientific community together with particle physics \citep{Knoedlseder2024}, while the path to carbon neutrality targets emissions limited to 2 \tCO{} per year per person by 2050. As scientists studying the habitability of other planets or the apparition of life, astronomers have an ethical responsibility to preserve our own planet. In order to raise awareness and be credible  with the public, astronomers must be beyond reproach when it comes to their environmental footprint (see the INSU 2025-2030 prospective studies on \emph{carbon footprint and transition\footnote{\url{https://prospective-aa.sciencesconf.org/data/rapport_synthese_groupeI2_transition_carbone_ecologique.pdf}}}) and to the social impact of the observatories in the territories where they are built  (see the INSU 2025-2030 prospective studies on \emph{telescope and territories, citizen sciences\footnote{\url{https://prospective-aa.sciencesconf.org/data/rapport_synthese_groupeI3_telescopes_territoires_astronomie_participative.pdf}}}).  While some astronomers stopped flying, others decided to use only archival data to avoid the race for larger and larger telescopes in space or on the ground. We explore here an alternative way, to branch out into the topic of light pollution, in an attempt to bridge the knowledge-action gap, that is, this huge disproportionality between how much astronomers know and how little they engage, as brilliantly explained in \citet{Dupont2024}. %


 

\subsection{Artificial light at night }

 Artificial light at night (hereafter ALAN) is one of the causes of biodiversity loss, alongside soil artificialisation and pesticides \cite[e.g.][for reviews]{Bennie2018,Sordello2022}. It is steadily increasing worldwide, with both a rise in sky brightness in affected areas and an increase in impacted surfaces, estimated between 2\% and 10\% per year \citep{Kyba2017,Kyba2023}. Astrophysicists were the first to sound the alarm in the late 1950s when major professional observatories were affected, followed by amateur astronomers advocating for the preservation of the sky and its cultural heritage. Flagstaff, Arizona was the first city to pass laws limiting the use of ALAN, and the International Astronomy Union gathered in Grenoble in 1976 for its XVI$^{th}$ general assembly recognized \emph{"the increasing levels of interference with astronomical observation resulting from artificial illumination of the night sky"} and urgently prompted \emph{"the responsible civil authorities to take action to preserve existing and planned observatories from such interferences"}. Today, we refer to the preservation of the night environment to encompass more globally the whole night-time ecosystem. 


\section{The transdisciplinary project of an observatory of the night environment }

\subsection{Birth of the project}

\begin{figure}[ht!]
 \centering
 \includegraphics[width=1.0\textwidth,clip]{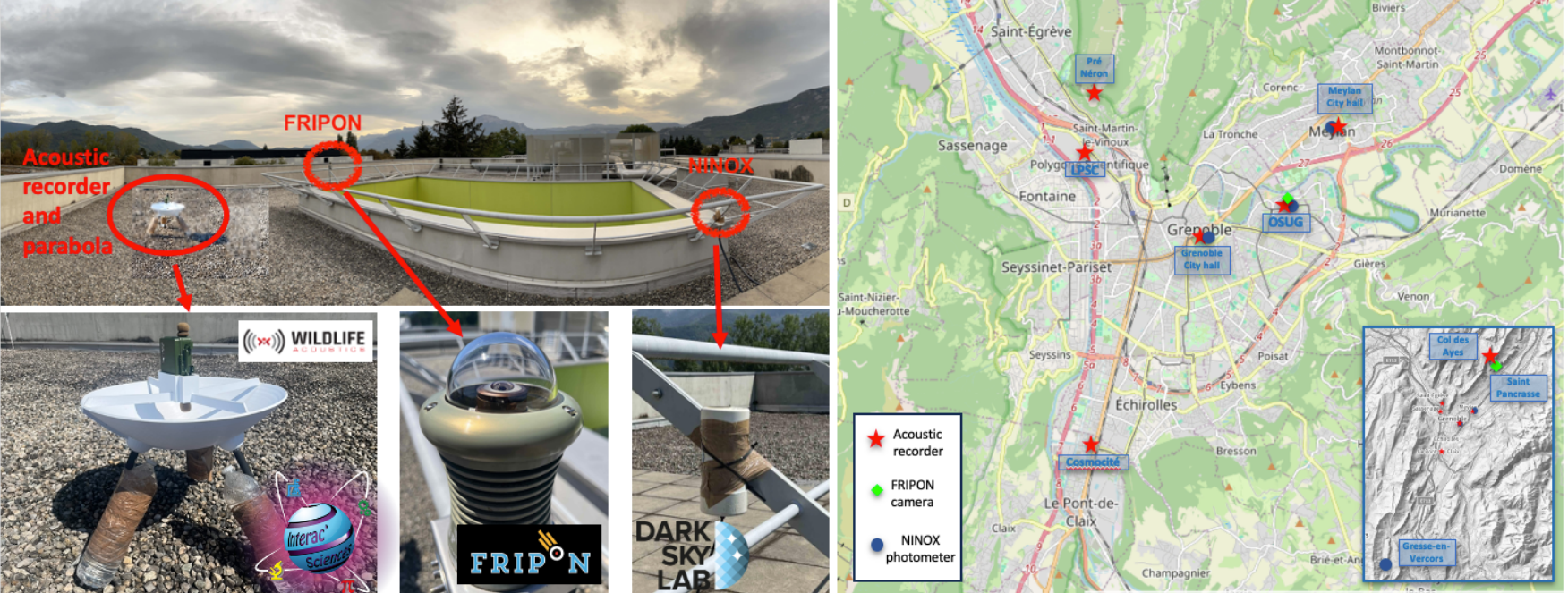}    
  \caption{{\bf Left:} The roof of OSUG equipped with the NINOX photometer, the FRIPON all-sky camera and two acoustic recorders, one being at the focus of a 40cm 3D-printed parabola {\bf Right:} Map of the Grenoble metropolitan area}
  \label{fig_map}
\end{figure}

Since 2014, astronomers from the the \emph{Institut de Plan\'etologie et d'Astrophysique de Grenoble} (IPAG) have been raising awareness on the issue of light pollution during the \emph{Jour de la Nuit}, an event \footnote{The \emph{Jour de la Nuit}, or \emph{Day of the Night}, is a national event coordinated by the association \emph{Agir pour l'environnement}} particularly observed in the Grenoble metropolitan area, where public lighting is partially turned off between 9pm and 11pm during onen night. Since 2021, we installed sensors on the roof of the \emph{Observatoire des Sciences de l'Univers de Grenoble} (OSUG, the federation of institutes including IPAG), see Fig. \ref{fig_map}, to measure light pollution locally, using the calibrated instrument NINOX, a photometer built by DarkSkyLab \footnote{\url{https://www.darkskylab.com}}. The goal was to directly measure the decrease in night sky brightness during the light extinction. We soon realised that the all-sky camera FRIPON (\citet{2020A&A...644A..53C} - Fireball Recovery and InterPlanetary Observation Network\footnote{\url{https://www.fripon.org}}) also installed on the same roof and delivering data since 2016, could be used to reanalyse archival images stored every 10min to extract the history of light pollution. Following this work, we established collaborations with the \emph{Ligue de Protection des Oiseaux} (LPO) to measure the impact of lighting on birds on the University Grenoble Alpes (UGA) campus, which was designated an \emph{LPO refuge\footnote{LPO refuges are spaces, often public parks and gardens, where the local authority is committed to biodiversity by creating appropriate facilities (e.g. nestboxes... ), raising awareness or monitoring species.}} since 2023, as well as at a site in downtown Grenoble. These actions allowed us to document the consequences of lighting choices in the Grenoble metropolitan area, interact with municipalities, and raise awareness among individuals about the impact of public and private lighting.

\subsection{Project goals, methods and partners}

The goal of the project is twofold. On the one hand, we want to extract the history of light pollution around Grenoble using archival data from FRIPON, to study the evolution of the lighting practices in the Grenoble metropolis. On the other hand, we want to study the impact of ALAN on ecosystems, particularly on migrating birds, a topic still poorly known. We designed this project at the local level of the Grenoble metropolitan area, but with the ambition to expand this pilot study to a regional or national level if successful. We sought support among the academic world but also among associations and municipalities, involving for instance the \emph{Laboratoire d'Ecologie Alpine} (LECA, also part of OSUG), the associations LPO and AurorAlpes, the company DarkSkyLab and the city council of several towns and the metropolis of Grenoble. We obtained funding from the LabEx OSUG and from  UGA and its program \emph{"Sciences en transition"}. FRIPON data have their own astrometric and photometric calibration \citep{2019A&A...627A..78J}, but is was important to compare them with Ninox data. To calibrate photometrically the FRIPON camera, we used the 1.5 year contemporaneous measurements with the NINOX photometer. We extracted from FRIPON images the median sky brightness in analog-to-digital-units in a cone around zenith with a diameter of $20^\circ$ (the full width at half-maximum of the field-of-view for the Unihedron SQM-LU inside NINOX) and selected cloudless nights without moon to calibrate these values against the night sky brightness measured by NINOX in \mpa \citep{Boribon2024}. To assess the impact of ALAN on migrating birds, we o equipped the roof of OSUG already hosting NINOX and FRIPON with an acoustic recorder monitoring nocturnal flight calls from birds from sunset to sun dawn. As some species can fly at high altitudes, we designed a 40cm parabola pointing towards zenith, such that one of the two microphones of the recorder is located at the focus of the parabola and receives amplified sounds from zenith, while the other microphone captures ambient sounds isotropically. Several additional sites were later equipped (see Fig. \ref{fig_map}).


\section{Results and conclusions}
 
\subsection{Measurements of the night sky brightness at the OSUG site}

\begin{figure}[ht!]
 \centering
 \includegraphics[width=1.0\textwidth,clip]{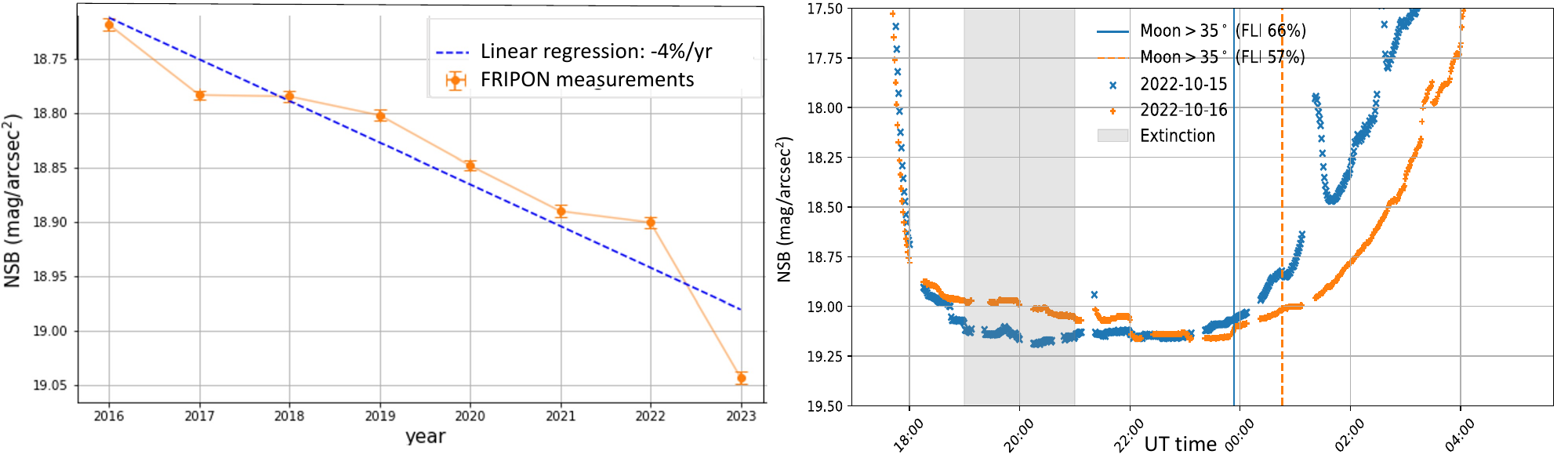}    
  \caption{{\bf Left:} History of night sky brightness from 2016 to 2023, as measured by FRIPON. {\bf Right:} Comparison between 2 nights: 15th October 2022 when a light extinction happened and the following night with similar weather conditions}
  \label{fig_ALAN_monitoring}
\end{figure}

\begin{figure}[ht!]
 \centering
 \includegraphics[width=1.0\textwidth,clip]{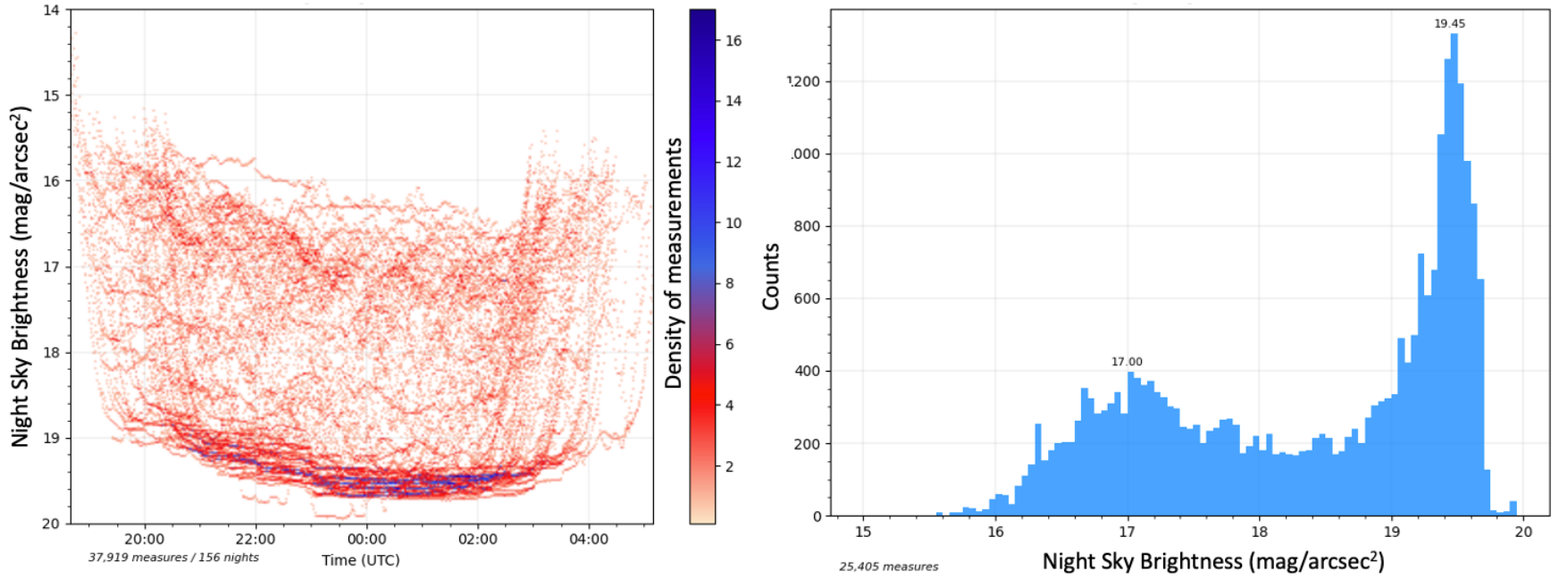}
  \caption{{NSB measured with NINOX for the OSUG site from April to September 2024. \bf Left:} Density plot combining measurements on moonless nights. {\bf Right:} Histogram of the measurements from the left plot excluding twilights.}
  \label{fig_ninox}
\end{figure}

The night sky brightness (NSB) as measured during cloudless and moonless astronomical nights decreased by 0.35 mag over 7 years, as shown in Fig. \ref{fig_ALAN_monitoring} (left), with a mean decrease of 4\% per year. This trend and the steeper decrease between 2022 and 2023 due to a wider practice of light extinction in the middle of the night, is also observed based on satellite data from the VIIRS DNB instrument. During the "Day of the night 2022" where a light extinction was implemented by some municipalities from the Grenoble metropolis, a decrease in NSB by 0.15 mag was observed (Fig. \ref{fig_ALAN_monitoring} right), corresponding to relative change by $-15\%$. Even though only some neighbourhoods and municipalities had agreed to switch off their public light on that day, this result highlights that urban lighting is a minor contribution to the overall level of light pollution which is dominated by some large private contributors such as industrial platforms, hospitals, hypermarket car parks, a conclusion also reached by \citet{Kyba2021} for the city of Tucson, USA.  When clouds are present, the level of NSB can increase by several orders of magnitudes, as visible in Fig. \ref{fig_ninox}. The denser areas shown in blue on the left plot indicate the typical NSB reached in the middle of the night in the absence of clouds and moons \cite[see][for details]{Deverchere2022}, and corresponds to the peak of the histogram shown on the right at $\sim19.5$ \mpa.

\subsection{Migration of nocturnal birds}

This is still too early to bring any conclusions regarding the study on the impact of ALAN on the nocturnal bird migration but we describe here the on-going work. A convolutional neural network, BirdNET, is used to detect and identify bird calls. A manual screening is currently required to validate the initial results and correct the algorithm inaccuracies. Given the difficulty of obtaining a number of individuals on the basis of sound recordings alone, the number of calls will be used as a proxy for the number of individuals. We hope to assess the daily and seasonal phenology, by species and compare that with the various environmental parameters (level of NSB, but also weather conditions such as the cloud cover, wind, rainfall in the air column above the recorder and on a larger scale in order to evaluate comparable days. The NSB will be the main parameter tested in order to assess the translations of migratory activity. Two main types of translation will be investigated: vertical and horizontal. Some species migrate at high altitude and could be attracted by the strong light sources of urban halos, and therefore greatly reduce their flight altitude. Other species, which migrate at low altitude, could change their horizontal flight path, which requires the deployment of a network of recorders around Grenoble and at less well-lit sites in the surrounding Alps. Given the presence of a network of FRIPON cameras throughout the region, a larger-scale deployment will be envisaged in order to obtain reliable data on horizontal translation, independent of topography and other environmental conditions that could influence these movements.

\bibliographystyle{aa}  
\bibliography{biblio} 

\newcommand{\noop}[1]{}
\begin{thebibliography}{12}
\expandafter\ifx\csname natexlab\endcsname\relax\def\natexlab#1{#1}\fi

\bibitem[{Bennie {et~al.}(2018)Bennie, Davies, Cruse, Bell, \& Gaston}]{Bennie2018}
Bennie, J., Davies, T.~W., Cruse, D., Bell, F., \& Gaston, K.~J. 2018, Journal of Applied Ecology, 55, 442

\bibitem[{Boribon(2024)}]{Boribon2024}
Boribon, M. 2024, Master's thesis, \url{https://dumas.ccsd.cnrs.fr/dumas-04713961}

\bibitem[{{Colas} {et~al.}(2020){Colas}, {Zanda}, {Bouley}, {Jeanne}, {Malgoyre}, {Birlan}, {Blanpain}, {Gattacceca}, {Jorda}, {Lecubin}, {Marmo}, {Rault}, {Vaubaillon}, {Vernazza}, {Yohia}, {Gardiol}, {Nedelcu}, {Poppe}, {Rowe}, {Forcier}, {Koschny}, {Trigo-Rodriguez}, {Lamy}, {Behrend}, {Ferri{\`e}re}, {Barghini}, {Buzzoni}, {Carbognani}, {Di Carlo}, {Di Martino}, {Knapic}, {Londero}, {Pratesi}, {Rasetti}, {Riva}, {Stirpe}, {Valsecchi}, {Volpicelli}, {Zorba}, {Coward}, {Drolshagen}, {Drolshagen}, {Hernandez}, {Jehin}, {Jobin}, {King}, {Nitschelm}, {Ott}, {Sanchez-Lavega}, {Toni}, {Abraham}, {Affaticati}, {Albani}, {Andreis}, {Andrieu}, {Anghel}, {Antaluca}, {Antier}, {App{\'e}r{\'e}}, {Armand}, {Ascione}, {Audureau}, {Auxepaules}, {Avoscan}, {Baba Aissa}, {Bacci}, {B{\v{a}}descu}, {Baldini}, {Baldo}, {Balestrero}, {Baratoux}, {Barbotin}, {Bardy}, {Basso}, {Bautista}, {Bayle}, {Beck}, {Bellitto}, {Belluso}, {Benna}, {Benammi}, {Beneteau}, {Benkhaldoun}, {Bergamini}, {Bernardi}, {Bertaina}, {Bessin}, {Betti},
  {Bettonvil}, {Bihel}, {Birnbaum}, {Blagoi}, {Blouri}, {Boac{\u{a}}}, {Boat{\v{a}}}, {Bobiet}, {Bonino}, {Boros}, {Bouchet}, {Borgeot}, {Bouchez}, {Boust}, {Boudon}, {Bouman}, {Bourget}, {Brandenburg}, {Bramond}, {Braun}, {Bussi}, {Cacault}, {Caillier}, {Calegaro}, {Camargo}, {Caminade}, {Campana}, {Campbell-Burns}, {Canal-Domingo}, {Carell}, {Carreau}, {Cascone}, {Cattaneo}, {Cauhape}, {Cavier}, {Celestin}, {Cellino}, {Champenois}, {Chennaoui Aoudjehane}, {Chevrier}, {Cholvy}, {Chomier}, {Christou}, {Cricchio}, {Coadou}, {Cocaign}, {Cochard}, {Cointin}, {Colombi}, {Colque Saavedra}, {Corp}, {Costa}, {Costard}, {Cottier}, {Cournoyer}, {Coustal}, {Cremonese}, {Cristea}, {Cuzon}, {D'Agostino}, {Daiffallah}, {D{\v{a}}nescu}, {Dardon}, {Dasse}, {Davadan}, {Debs}, {Defaix}, {Deleflie}, {D'Elia}, {De Luca}, {De Maria}, {Deverch{\`e}re}, {Devillepoix}, {Dias}, {Di Dato}, {Di Luca}, {Dominici}, {Drouard}, {Dumont}, {Dupouy}, {Duvignac}, {Egal}, {Erasmus}, {Esseiva}, {Ebel}, {Eisengarten}, {Federici}, {Feral},
  {Ferrant}, {Ferreol}, {Finitzer}, {Foucault}, {Francois}, {Fr{\^\i}ncu}, {Froger}, {Gaborit}, {Gagliarducci}, {Galard}, {Gardavot}, {Garmier}, {Garnung}, {Gautier}, {Gendre}, {Gerard}, {Gerardi}, {Godet}, {Grandchamps}, {Grouiez}, {Groult}, {Guidetti}, {Giuli}, {Hello}, {Henry}, {Herbreteau}, {Herpin}, {Hewins}, {Hillairet}, {Horak}, {Hueso}, {Huet}, {Huet}, {Hyaum{\'e}}, {Interrante}, {Isselin}, {Jeangeorges}, {Janeux}, {Jeanneret}, {Jobse}, {Jouin}, {Jouvard}, {Joy}, {Julien}, {Kacerek}, {Kaire}, {Kempf}, {Koschny}, {Krier}, {Kwon}, {Lacassagne}, {Lachat}, {Lagain}, {Laisn{\'e}}, {Lanchares}, {Laskar}, {Lazzarin}, {Leblanc}, {Lebreton}, {Lecomte}, {Le D{\^u}}, {Lelong}, {Lera}, {Leoni}, {Le-Pichon}, {Le-Poupon}, {Leroy}, {Leto}, {Levansuu}, {Lewin}, {Lienard}, {Licchelli}, {Locatelli}, {Loehle}, {Loizeau}, {Luciani}, {Maignan}, {Manca}, {Mancuso}, {Mandon}, {Mangold}, {Mannucci}, {Maquet}, {Marant}, {Marchal}, {Marin}, {Martin-Brisset}, {Martin}, {Mathieu}, {Maury}, {Mespoulet}, {Meyer}, {Meyer}, {Meza},
  {Moggi Cecchi}, {Moiroud}, {Millan}, {Montesarchio}, {Misiano}, {Molinari}, {Molau}, {Monari}, {Monflier}, {Monkos}, {Montemaggi}, {Monti}, {Moreau}, {Morin}, {Mourgues}, {Mousis}, {Nablanc}, {Nastasi}, {Niac{\c{s}}u}, {Notez}, {Ory}, {Pace}, {Paganelli}, {Pagola}, {Pajuelo}, {Palaci{\'a}n}, {Pallier}, {Paraschiv}, {Pardini}, {Pavone}, {Pavy}, {Payen}, {Pegoraro}, {Pe{\~n}a-Asensio}, {Perez}, {P{\'e}rez-Hoyos}, {Perlerin}, {Peyrot}, {Peth}, {Pic}, {Pietronave}, {Pilger}, {Piquel}, {Pisanu}, {Poppe}, {Portois}, {Prezeau}, {Pugno}, {Quantin}, {Quitt{\'e}}, {Rambaux}, {Ravier}, {Repetti}, {Ribas}, {Richard}, {Richard}, {Rigoni}, {Rivet}, {Rizzi}, {Rochain}, {Rojas}, {Romeo}, {Rotaru}, {Rotger}, {Rougier}, {Rousselot}, {Rousset}, {Rousseu}, {Rubiera}, {Rudawska}, {Rudelle}, {Ruguet}, {Russo}, {Sales}, {Sauzereau}, {Salvati}, {Schieffer}, {Schreiner}, {Scribano}, {Selvestrel}, {Serra}, {Shengold}, {Shuttleworth}, {Smareglia}, {Sohy}, {Soldi}, {Stanga}, {Steinhausser}, {Strafella}, {Sylla Mbaye}, {Smedley},
  {Tagger}, {Tanga}, {Taricco}, {Teng}, {Tercu}, {Thizy}, {Thomas}, {Tombelli}, {Trangosi}, {Tregon}, {Trivero}, {Tukkers}, {Turcu}, {Umbriaco}, {Unda-Sanzana}, {Vairetti}, {Valenzuela}, {Valente}, {Varennes}, {Vauclair}, {Vergne}, {Verlinden}, {Vidal-Alaiz}, {Vieira-Martins}, {Viel}, {V{\^\i}ntdevar{\v{a}}}, {Vinogradoff}, {Volpini}, {Wendling}, {Wilhelm}, {Wohlgemuth}, {Yanguas}, {Zagarella}, \& {Zollo}}]{2020A&A...644A..53C}
{Colas}, F., {Zanda}, B., {Bouley}, S., {et~al.} 2020, \aap, 644, A53

\bibitem[{Deverchère {et~al.}(2022)Deverchère, Vauclair, Bosch, Moulherat, \& Cornuau}]{Deverchere2022}
Deverchère, P., Vauclair, S., Bosch, G., Moulherat, S., \& Cornuau, J.~H. 2022, Scientific Reports, 12, 17050, nature Publishing Group

\bibitem[{Dupont {et~al.}(2024)Dupont, Jacob, \& Philippe}]{Dupont2024}
Dupont, L., Jacob, S., \& Philippe, H. 2024, Nature Ecology \& Evolution

\bibitem[{{Jeanne} {et~al.}(2019){Jeanne}, {Colas}, {Zanda}, {Birlan}, {Vaubaillon}, {Bouley}, {Vernazza}, {Jorda}, {Gattacceca}, {Rault}, {Carbognani}, {Gardiol}, {Lamy}, {Baratoux}, {Blanpain}, {Malgoyre}, {Lecubin}, {Marmo}, \& {Hewins}}]{2019A&A...627A..78J}
{Jeanne}, S., {Colas}, F., {Zanda}, B., {et~al.} 2019, \aap, 627, A78

\bibitem[{{Kn{\"o}dlseder} {et~al.}(2022){Kn{\"o}dlseder}, {Brau-Nogu{\'e}}, {Coriat}, {Garnier}, {Hughes}, {Martin}, \& {Tibaldo}}]{Knoedlseder2022}
{Kn{\"o}dlseder}, J., {Brau-Nogu{\'e}}, S., {Coriat}, M., {et~al.} 2022, Nature Astronomy, 6, 503

\bibitem[{{Kn{\"o}dlseder} {et~al.}(2024){Kn{\"o}dlseder}, {Coriat}, {Garnier}, \& {Hughes}}]{Knoedlseder2024}
{Kn{\"o}dlseder}, J., {Coriat}, M., {Garnier}, P., \& {Hughes}, A. 2024, Nature Astronomy

\bibitem[{Kyba {et~al.}(2021)Kyba, Ruby, Kuechly, Kinzey, Miller, Sanders, Barentine, Kleinodt, \& Espey}]{Kyba2021}
Kyba, C., Ruby, A., Kuechly, H., {et~al.} 2021, Lighting Research \& Technology, 53, 189, sAGE Publications Ltd STM

\bibitem[{Kyba {et~al.}(2023)Kyba, Altintas, Walker, \& Newhouse}]{Kyba2023}
Kyba, C. C.~M., Altintas, Y.~O., Walker, C.~E., \& Newhouse, M. 2023, Science, 379, 265, american Association for the Advancement of Science

\bibitem[{Kyba {et~al.}(2017)Kyba, Kuester, Sánchez~de Miguel, Baugh, Jechow, Hölker, Bennie, Elvidge, Gaston, \& Guanter}]{Kyba2017}
Kyba, C. C.~M., Kuester, T., Sánchez~de Miguel, A., {et~al.} 2017, Science Advances, 3, e1701528, american Association for the Advancement of Science

\bibitem[{Sordello {et~al.}(2022)Sordello, Busson, Cornuau, Deverchère, Faure, Guetté, Hölker, Kerbiriou, Lengagne, Le~Viol, Longcore, Moeschler, Ranzoni, Ray, Reyjol, Roulet, Schroer, Secondi, Valet, Vanpeene, \& Vauclair}]{Sordello2022}
Sordello, R., Busson, S., Cornuau, J.~H., {et~al.} 2022, Landscape and Urban Planning, 219, 104332

\end{thebibliography}

\end{document}